\documentclass[journal=jacsat,manuscript=article]{achemso}

\usepackage{chemformula} 
\usepackage[T1]{fontenc} 
\usepackage{mathalpha}
\usepackage{amssymb}
\usepackage{amsmath}
\usepackage[version=4]{mhchem}
\usepackage{graphicx}
\usepackage{float}
\usepackage{rotating}
\usepackage{placeins}
\usepackage[colorlinks=true,citecolor=blue]{hyperref}
\usepackage{color,soul}
\let\oldAA\AA
\renewcommand{\AA}{\text{\normalfont\oldAA}}
\author{Y.T. Singh}
\affiliation[North-Eastern Hill University]
{Department of Physics, North-Eastern Hill University, Shillong, Meghalaya, 793022, India}
\alsoaffiliation[Pachhunga University College]
{Physical Sciences Research Center (PSRC),  Department of Physics, Pachhunga University College, Mizoram University, Aizawl, 796001, Mizoram, India}
\email{yumnamthakur99@gmail.com}
\author{P. K. Patra}
\affiliation[North-Eastern Hill University]
{Department of Physics, North-Eastern Hill University, Shillong, 793022, Meghalaya, India}
\email{pkpatranehu@gmail.com}
\author{D. P. Rai}
\affiliation[Pachhunga University College]
{Physical Sciences Research Center (PSRC),  Department of Physics, Pachhunga University College, Mizoram University, Aizawl, 796001, Mizoram, India}
\email{*Corresponding  Author:dibya@pucollege.edu.in}

\title[An \textsf{achemso} demo]
  {Electronic and mechanical properties of Nitrogen doped (6,1)  single walled carbon nanotube (SWCNT)from first-principles DFT and Molecular Dynamics approach}


\begin{document}







\begin{abstract}
 In this paper we have analysed the electronic and mechanical properties of Nitrogen(N) doped (6,1) SWCNTs based on first-principles and Molecular dynamic (MD) simulation. A schematic N-doping on SWCNT was performed along zigzag(zz) and armchair(ac) direction. Armchair doping is considered parallel to tube axis while zigzag is along cross-section perpendicular to tube axis. In doping pattern (both zz and ac) we have observed the variation in electronic properties for even number of N-doping and odd number of N-doping. To study the mechanical properties we have adopted $ab-initio$ MD-simulations. We report the dependent of tensile response of the tube on the dopant concentration and doping pattern. Single N-doped system shows enhanced tensile stress by 55\% as compared to the pristine SWCNT. While the variation of young's modulus for all N-doped systems are almost invariant.
\end{abstract}

\section{Introduction}
Since the
A multiwall carbon nanotube was first discovered by Iijima et al.\cite{Iijima1991}. Carbon nanotubes has become a promising material in the field of nano electronics due to their extraordinary electronic \cite{Fischer1999,Issi1995}, mechanical\cite{Mawphlang2020,Zhu2016} and optical properties\cite{Mizuno2009}\cite{Filho2004}. Earlier studies reported the dependence of the intrinsic electronic, mechanical and optical properties of a single walled carbon nanotube (SWCNT) to its  chirality and diameter based on their fabrication processes\cite{Sinnott2010,X2009}. Synthesis of  particular SWCNT with desire properties is very challenging. Many theoretical and experimental investigations have reported the modification of the electronic properties of SWCNTs by generating defect \cite{Jonuarti2017,Algharagholy2019}, doping with other elements \cite{Liu2020,Algharagholy2019} and applying stress\cite{Coppola2011,Heyd1997}. Modification by doping has been a promising method in the field of technological applications as their Fermi level can be easily manipulated. From first principle study, Bashir et al. \cite{UdDinBhat2016} theoretically investigated the effect of the dopant concentration on the electronic properties of N-doped (8,0) SWCNT. The authors reported that the semiconducting (8,0) SWCNT converted into metallic nanotube with increase in concentration of dopant.  Experimental investigation on SWCNT doped with polyethyleneimine reports enhancement in electrical conductivity by 42.2\% in comparison with undoped SWCNT \cite{MonikaRdest2020}. Similar result was reported by Azam et al., from the theoretical investigation (DFT) on the electrical conductivity of Si-doped SWCNT \cite{Azam2017}. Various scientific literature reported that CNTs doped with different elements posses unique sensitivity and selectivity for different gases which make them a promising materials for gas sensing device\cite{Zhang2009a, Mittal2014, Zaporotskova2016}.\par
Theoretical and experimental studies have predicted the effective Young's modulus value for SWCNTs is $\sim$1 TPa \cite{Hsieh2006, Wu2008}. But the ideal tensile strength and the critical strain value reported from the experimental investigations was lower than than the theoretical values\cite{Takakura2019, Yang2016}. Several researchers suggested that such lowering in experimental result  is due to existence of structural defects on the tube. Lin Yang et al. \cite{Yang2016}, from molecular dynamics study reported the reduction in ideal strength by 60\% for CNTs with common types of defects. Similar result was also reported by Liyan Zhu et al.\cite{Zhu2016}. Belytschko et al. \cite{Belytschko2002} from atomistic simulations report the dependence of the mechanical strength of CNTs to their chemical structure which was supported experimentally by Takakura et al.\cite{Takakura2019}. Fakhrabadi et al.\cite{Title2021}, from Molecular dynamic study reported the reduction in elastic property of (10,10) SWCNT when doped with nitrogen. Recently, Choi et al.\cite{Jung2020} theoretically studied the effect of doping on the mechanical properties of CNTs considering three different nitrogen doping groups (Quaternary, pyridinic, and pyrrolic), and reported the variation in mechanical properties.\par
In this work we report the systematic information about the variation of electronic and mechanical properties of SWCNT via doping of N-atoms in different directions (zz \& ac). We may look up this work as the first of its kind in calculating the doping-direction and doping-concentration dependent physical properties of SWCNT. We have adopted a newly developed first-principles based DFT-1/2 approach for all electronic calculations. For the mechanical properties, we have performed a series of  molecular dynamic simulations using NPT Martyna Tobias Klein algorithm \cite{Martyna1998} as implemented in VNL-ATK software pakage\cite{Smidstrup2020}.

\begin{figure}[H]
	\centering
	\includegraphics[width=10.0cm, height=3.50cm]{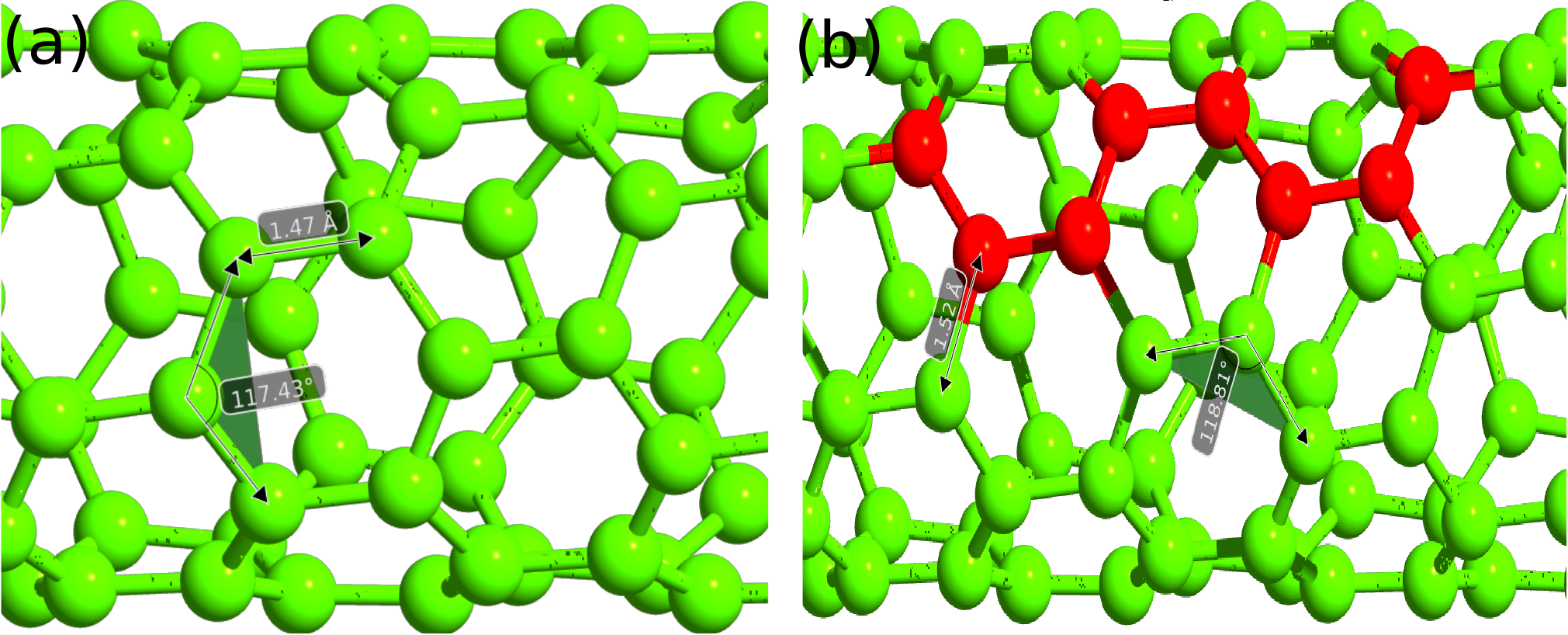}]
	\includegraphics[width=5.0cm, height=3.50cm]{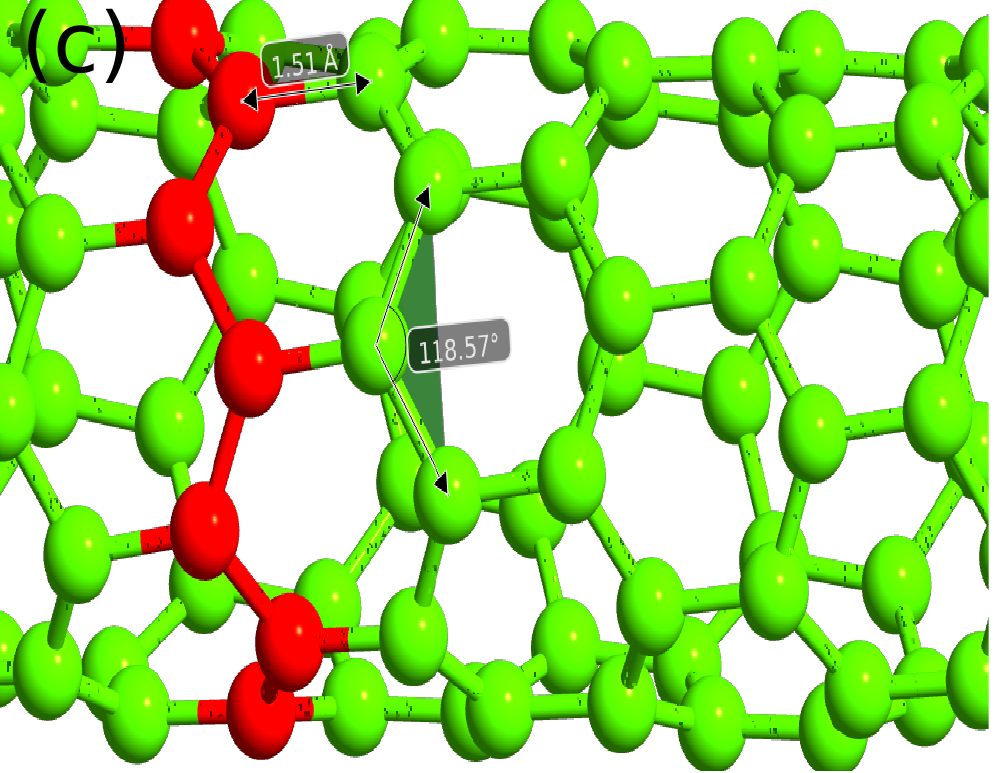}]
		\caption{\label{fig:CNT}. (a) Pristine (6,1) SWCNT , (b) arm chair doping system of (6,1) SWCNT, (c)   zig-zag doping system of (6,1) SWCNT }
	\end{figure}

\section{Computational detail}
  The carbon atoms having C-C bond-length of 1.45 \AA on the surface of the (6,1) SWCNT are periodically substituted by Nitrogen atoms. The periodicity of N atoms are considered in two different ways for our investigation. The two different pattern of dopings are; doping along the axis of the tube (arm-chair doping) and doping along curvature of the tube (zigzag structure) [see figure \ref{fig:CNT}].  All  the structures were geometrically optimized using Force field potential set "Teroff BNC 2000" \cite{K2000} until the forces reaches lower than 0.001 ev/\AA. To evaluate the  stability of the systems, formation energy is also calculated using Eq.\ref{1} \cite{Fujimoto2015}
  
  \begin{equation} \label{1}
  E_{form} = E_{t} - n_ {c}\times{E_{c}}	- n_{N}\times{E_{N}}	
  \end{equation}
  
  where $E_{t}$ is the total energy of  the system, $E_{c}$ is the energy of a carbon atom, $E_{N}$ is the energy of a nitrogen atom, $n_{c}$ and $n_{N}$ are the number of carbon atoms and the number of nitrogen atoms in the system respectively. 
  \par The electronic properties were calculated by using generalized DFT-1/2\cite{Ferreira2008}. DFT-1/2 is  a new type of approximation based on the Slater's half occupation technique for correcting the self-interaction error in local and semi-local exchange-correlation functional. The modified KS potential is $V_{mod-KS}(r)$=$V_{KS}$(r)-$V_{S}(r)$, where $V_{KS}$(r) is a standard DFT potential and $V_{S}(r)$ is a self-energy potential \cite{Ferreira2011,Yuan2018}
   We have also investigated the mechanical properties using MD-simulation technique. MD is a developed to characterized the effective properties at nanoscale. In our investigation the Teroff BNC 2000 classical potential\cite{K2000} is adopted as potential function for two C-N atoms. Setting the reservoir temperature at 300K and drawing the initial velocity of the atoms from the Maxwell Boltzmann distribution at 300K a series of NPT Molecular dynamic simulation is performed with total time  steps of  40000 fs  for each system within NPT Martyna Tobias Klein algorithm\cite{Martyna1998}. In each 1000 time step a strain value of 0.01 is applied and the corresponding stress value is calculated which is given by Eq.\ref{2} \cite{Zang2009a,Zhu2007}
  
  https://www.overleaf.com/project/6109da5d3e14b281a5686dac
  \begin{equation}\label{2}
  \sigma^{\alpha \beta}=1/V[-\Sigma_{i}m_{i}v^{\alpha}_{i}v^{\beta}_{i}+1/2\Sigma_{i}\Sigma_{j\not=i} F^{\alpha}_{ij}r^{\beta}_{ij}]
  \end{equation}
  
  here $m_{i}$  and $v_{i}$ are the mass and velocity of the of the i atom, $F_{ij}$ is the force between  $i$ and  $j$ atoms, $\alpha$ and $\beta$ are the Cartesian components, $V$ is the total volume occupied by all the atoms and $r^{\beta}_{ij}$ is the projection of the inter-atomic distance along the $\alpha$ coordinate.
  
\section{Result and discussion} 
\subsection{Electronic properties}
\begin{figure}[H]
	\centering
	\includegraphics[width=15.0cm, height=7.0cm]{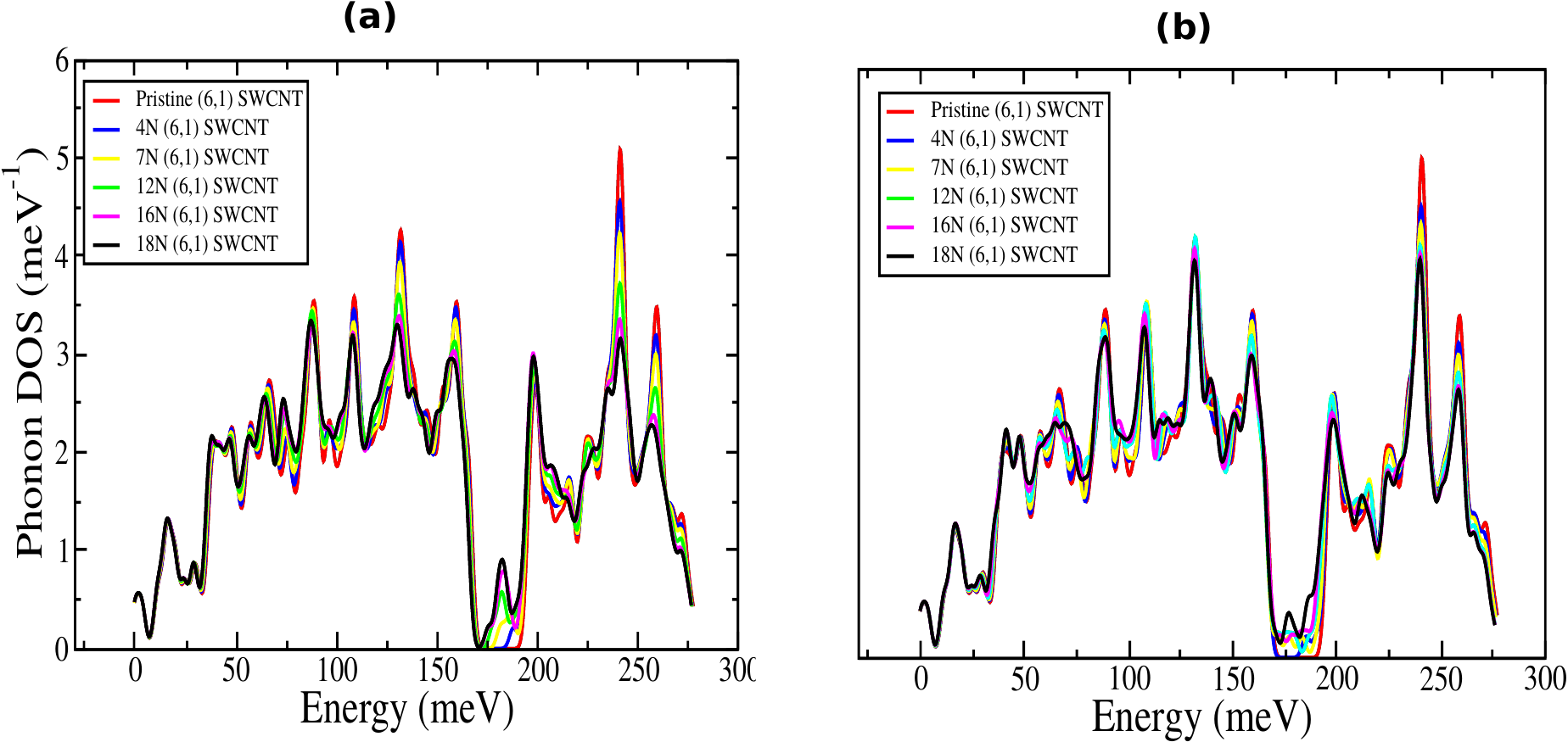}
	\caption{\label{fig:pdos}. Phonon DOS for (a) arm-chair doping systems , (b)   zig-zag doping systems }
\end{figure}
\begin{figure}[H]
	\centering
	\includegraphics[width=8.0cm, height=8.0cm]{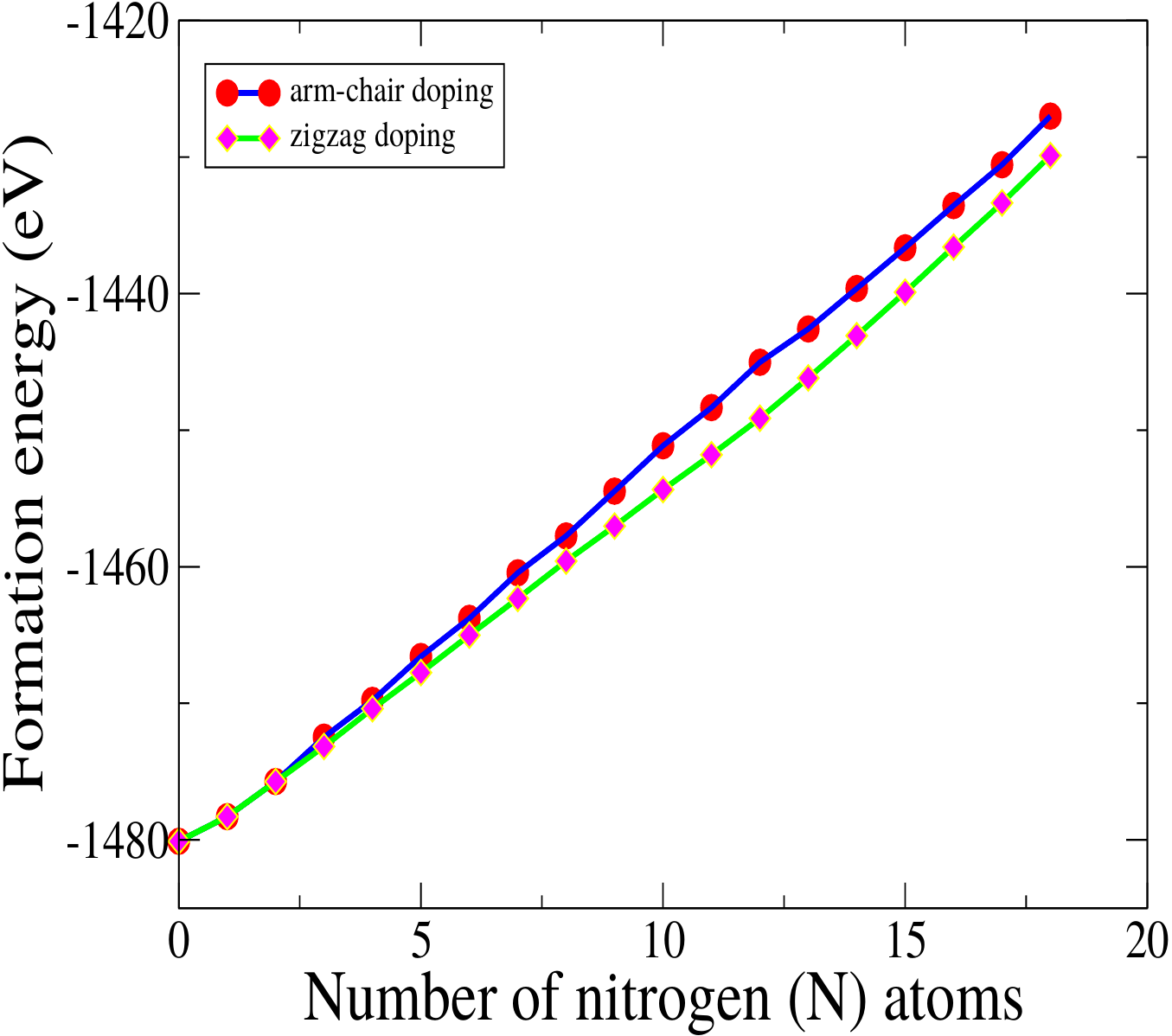}
	\caption{\label{fig:formation}.Formation energy and number of doped N atoms graph .}
\end{figure}

To analyse the thermal stability of the systems, we have calculated the phonon density of states (Phonon DOS) for all doped/undoped optimized systems. Figure \ref{fig:pdos} elucidate  the calculated results. No negative phonon DOS has been observed in our systems which indicate all the doped systems are thermally stable. For energy stability analysis of the systems, the formation energy for each system is calculated. The calculated results are shown in figure \ref{fig:formation}. All the systems show negative values of formation energies and hence the structural stability has been confirmed. In both arm-chair doping and zigzag doping, the stability decreases with increase in doping concentration with highest negative formation energy value of -1480.10318 eV has been reported for pristine CNT. Similar result was also reported by Jonuarti et al. \cite{Title2021} In comparing the formation energies of the systems for same doping concentration in two different pattern zz and ac, zz doping   exhibit the higher negative values of formation energy as compared  ac doping. 
\begin{figure}[H]
	\centering		
	\includegraphics[width=10.0cm, height=15.0cm]{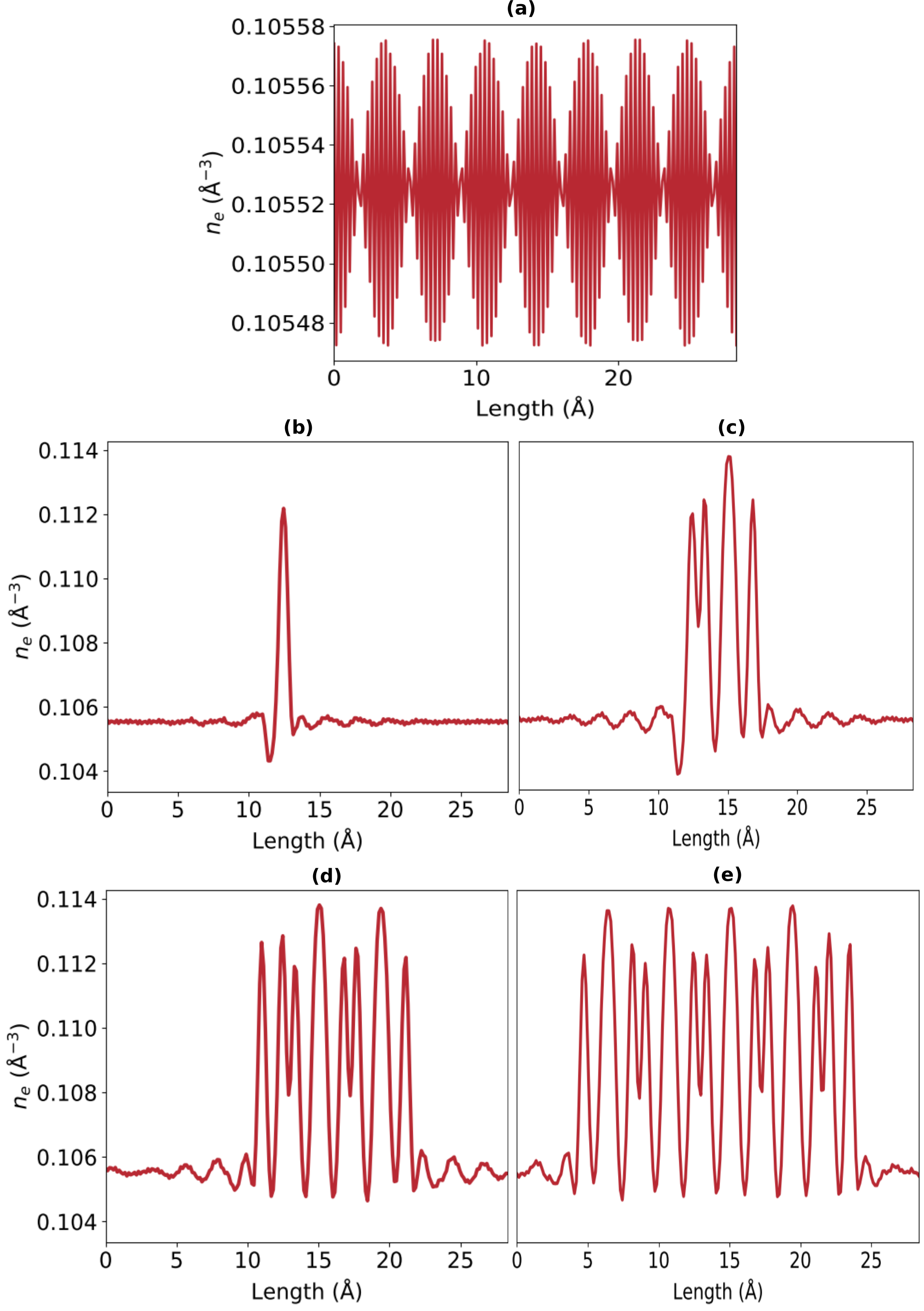}
	\caption{\label{fig:den1}.Electron density along the tube length for (a) Pristine (6,1) SWCNT, (b) 1N system, (c) 5N system, (d) 10N system, (e) 18N system with arm-chair doping}
\end{figure}
\begin{figure}[H]
	\centering
	\includegraphics[width=10.0cm, height=15.0cm]{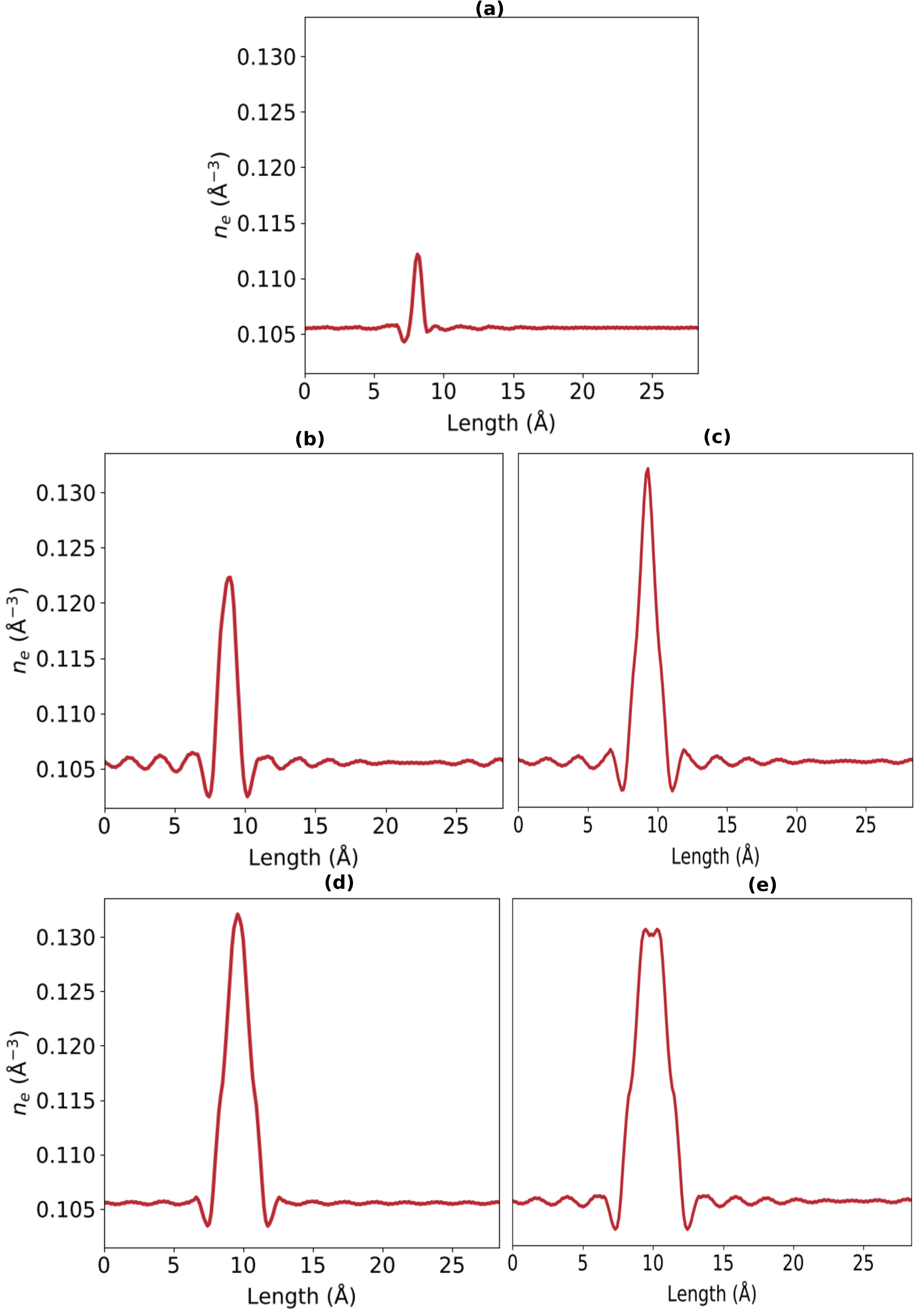}
	\caption{\label{fig:den2}. Electron density along the tube length for  (a) 1N system, (b) 5N system, (c) 10N system, (d) 13N system, (e) 18N system with zigzag doping}
\end{figure}

\begin{figure}[H]
	\centering		
	\includegraphics[width=10.0cm, height=15.0cm]{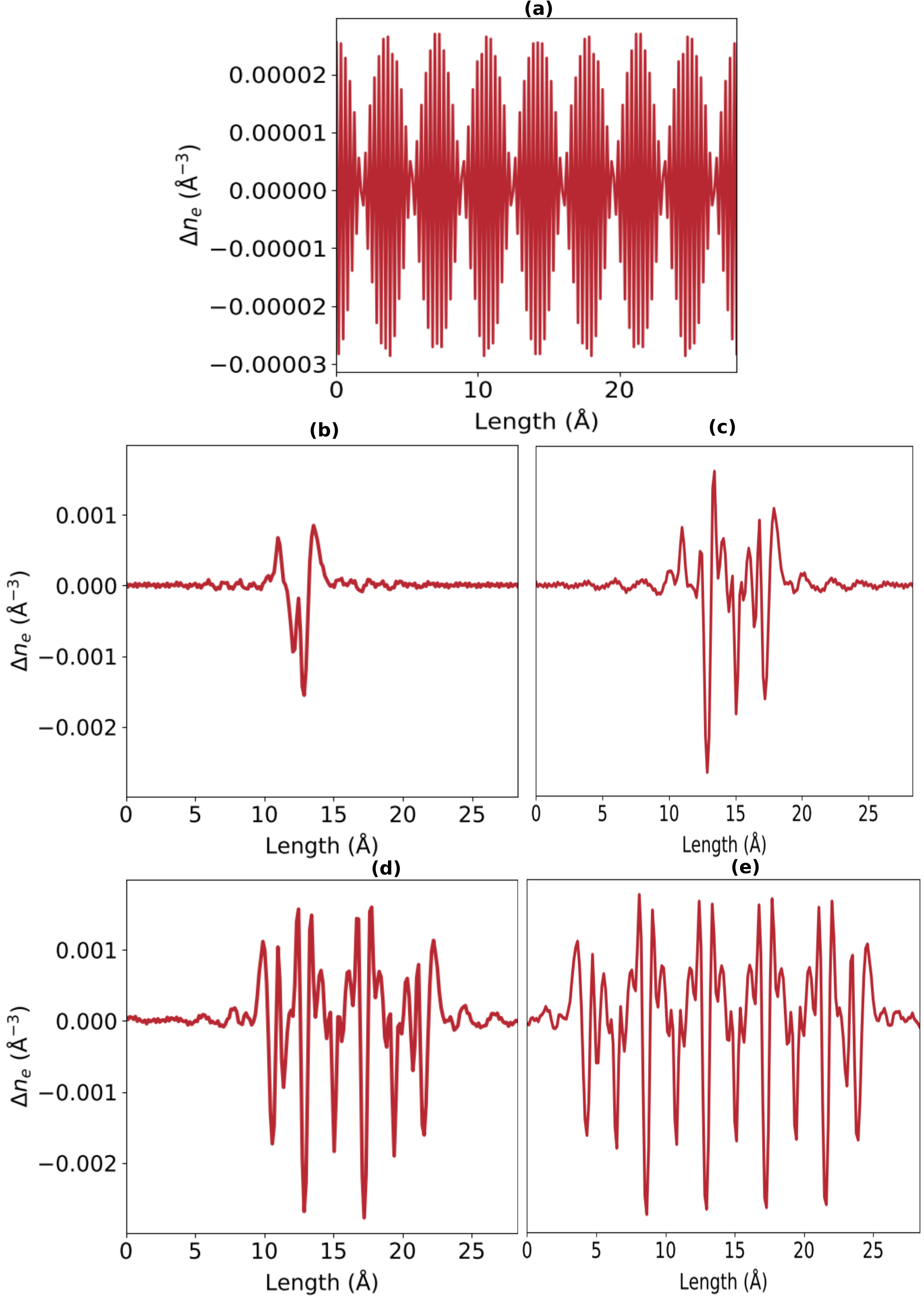}
	\caption{\label{fig:diff1}.Electron difference density along the tube length for (a) Pristine (6,1) SWCNT, (b) 1N system, (c) 5N system, (d) 10N system, (e) 18N system with arm-chair doping }
\end{figure}
\begin{figure}[H]
	\centering
	\includegraphics[width=10.0cm, height=15.0cm]{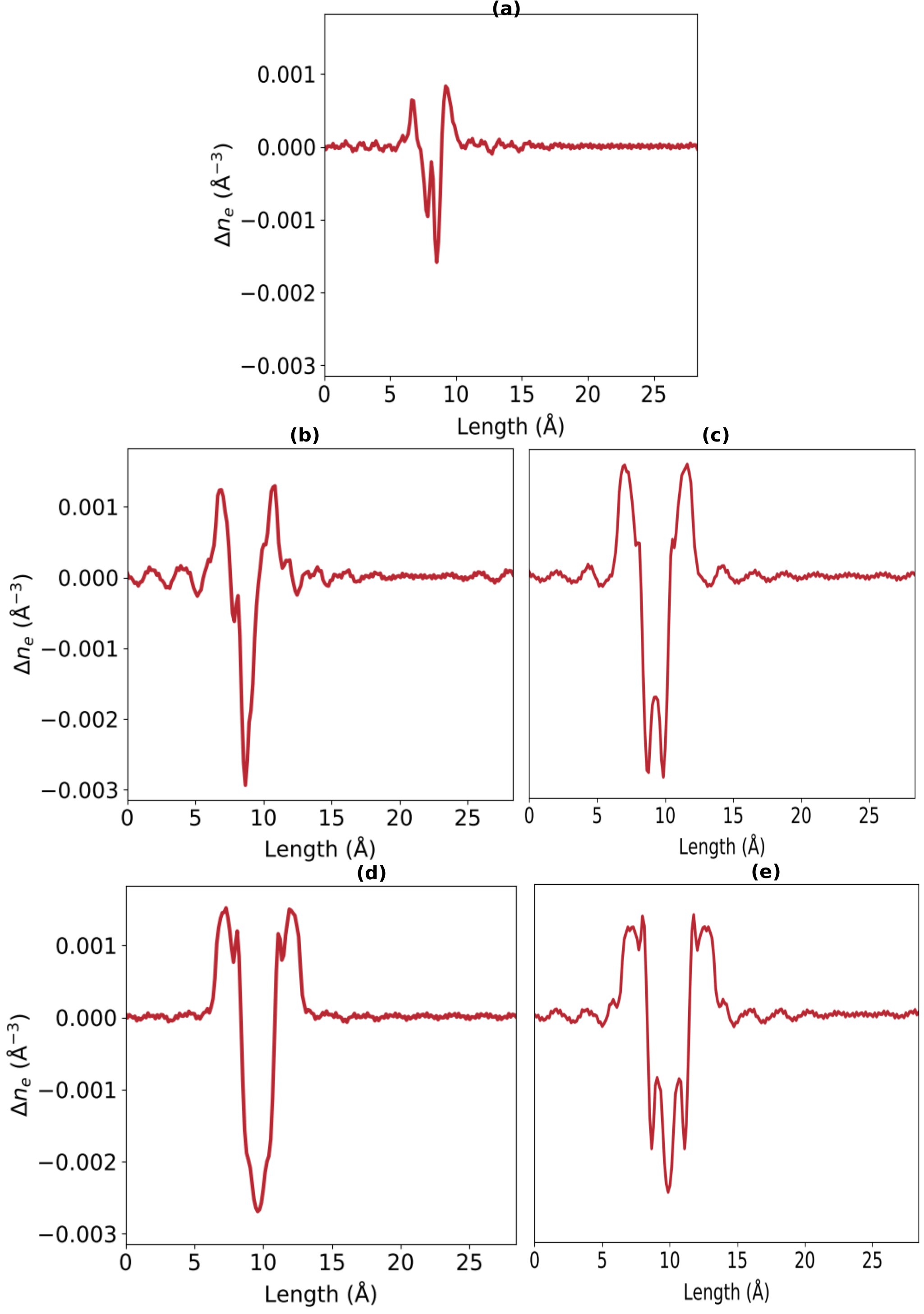}
	\caption{\label{fig:diff2}.Electron difference density along the tube length for  (a) 1N system, (b) 5N system, (c) 10N system, (d) 13N system, (e) 18N system with zigzag doping }
\end{figure}

As reported in our previous work pristine (6,1) SWCNT is a low band gap semiconductor, whose energy bandgap value is 0.471245 eV \cite{Rai2020}. When the carbon atoms on the surface of the SWCNT are replaced by the N ones, the bond lengths as well as the bond angles of the tube are changed [see figure \ref{fig:CNT}]. Such changes alter the tube structure and hence the electronic properties\cite{Bahari2017}. Figure (\ref{fig:den1} and \ref{fig:den2}) shows the electron density profile while figure (\ref{fig:diff1} and \ref{fig:diff2}) shows the difference of electron density profile of the  systems.  Higher electron density is observed at the sites of N doping. The electron density profile of the SWCNT changes significantly with doping concentration as well as doping pattern. This proves that the doping concentration and doping pattern  plays an important role in modulating the  electronic properties of the tube. 

\begin{figure}[H]
	\centering
	\includegraphics[width=10.0cm, height=15.0cm]{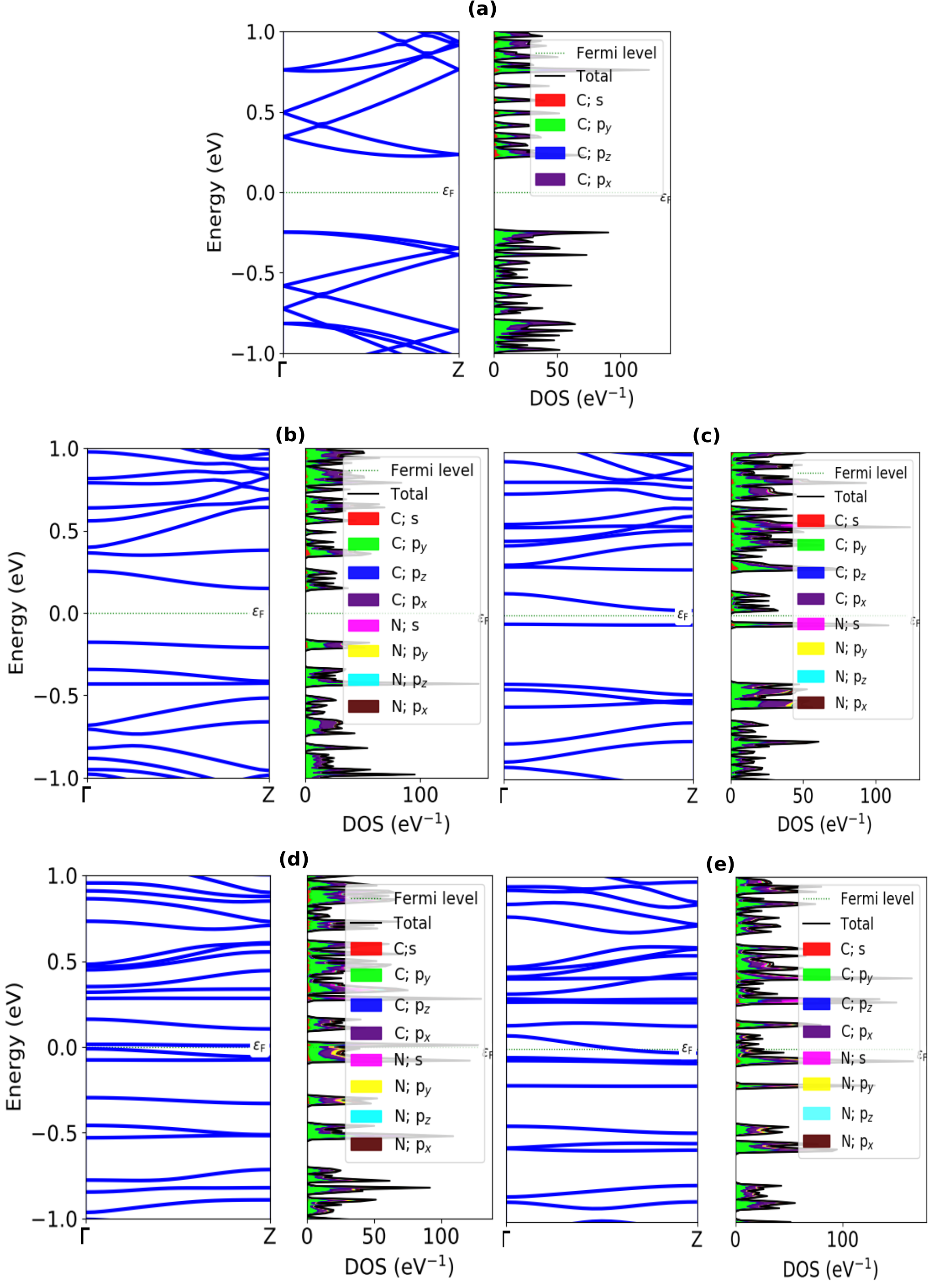}
	\caption{\label{fig:Band1}.Band structures and PDOS for (a) Pristine (6,1) SWCNT, (b) 2N system, (c) 6N system, (d) 10N system, (e) 12N system with arm-chair doping }
\end{figure}
\begin{figure}[H]
	\centering
	\includegraphics[width=10.0cm, height=15.0cm]{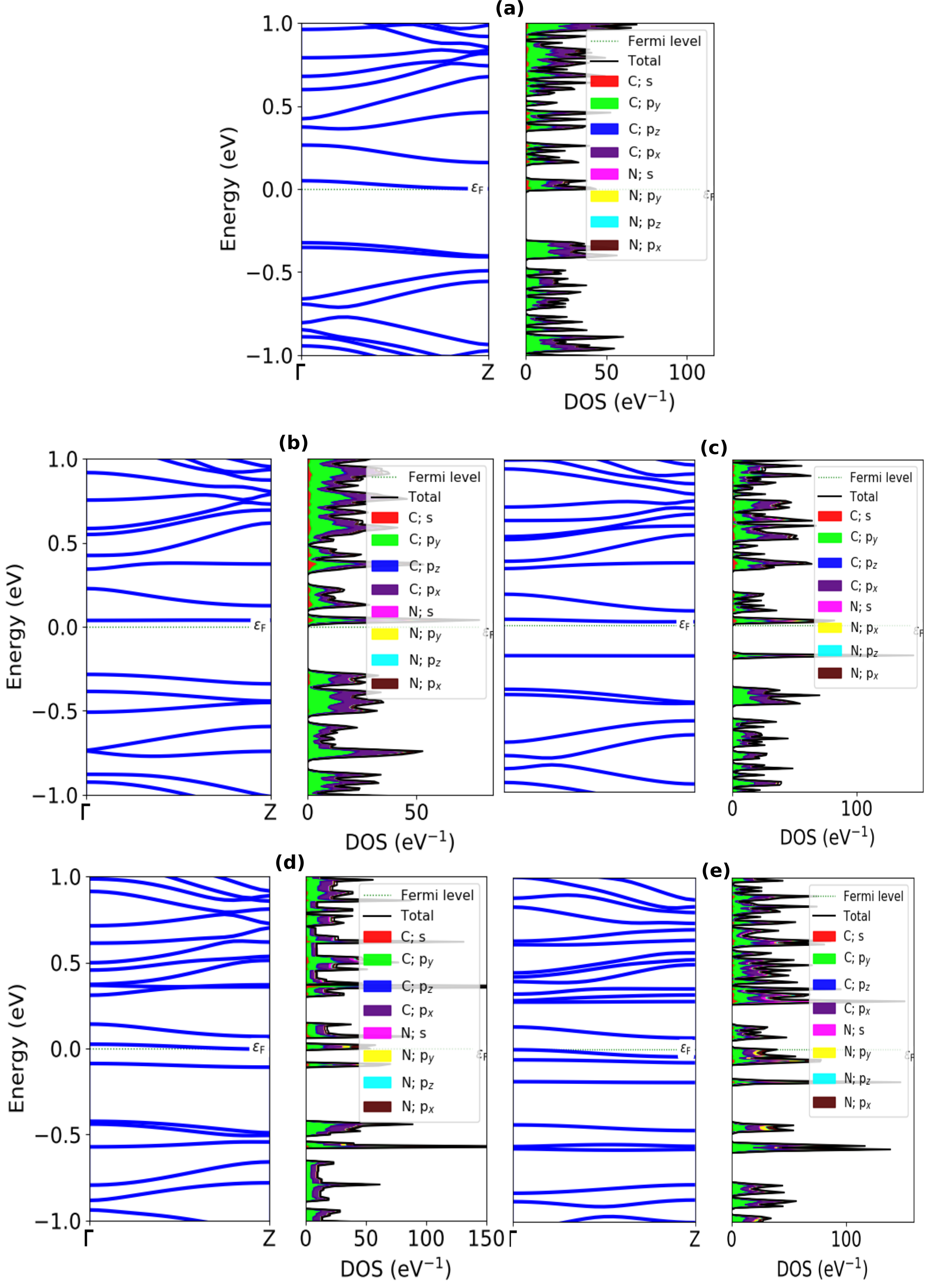}\\
	\caption{\label{fig:Band2}.Band structures and PDOS for (a) 1N system, (b) 3N system, (c) 5N system, (d) 7N system, (e) 11N system with arm-chair doping }
\end{figure}
\begin{figure}[H]
	\centering		
	\includegraphics[width=10.0cm, height=15.0cm]{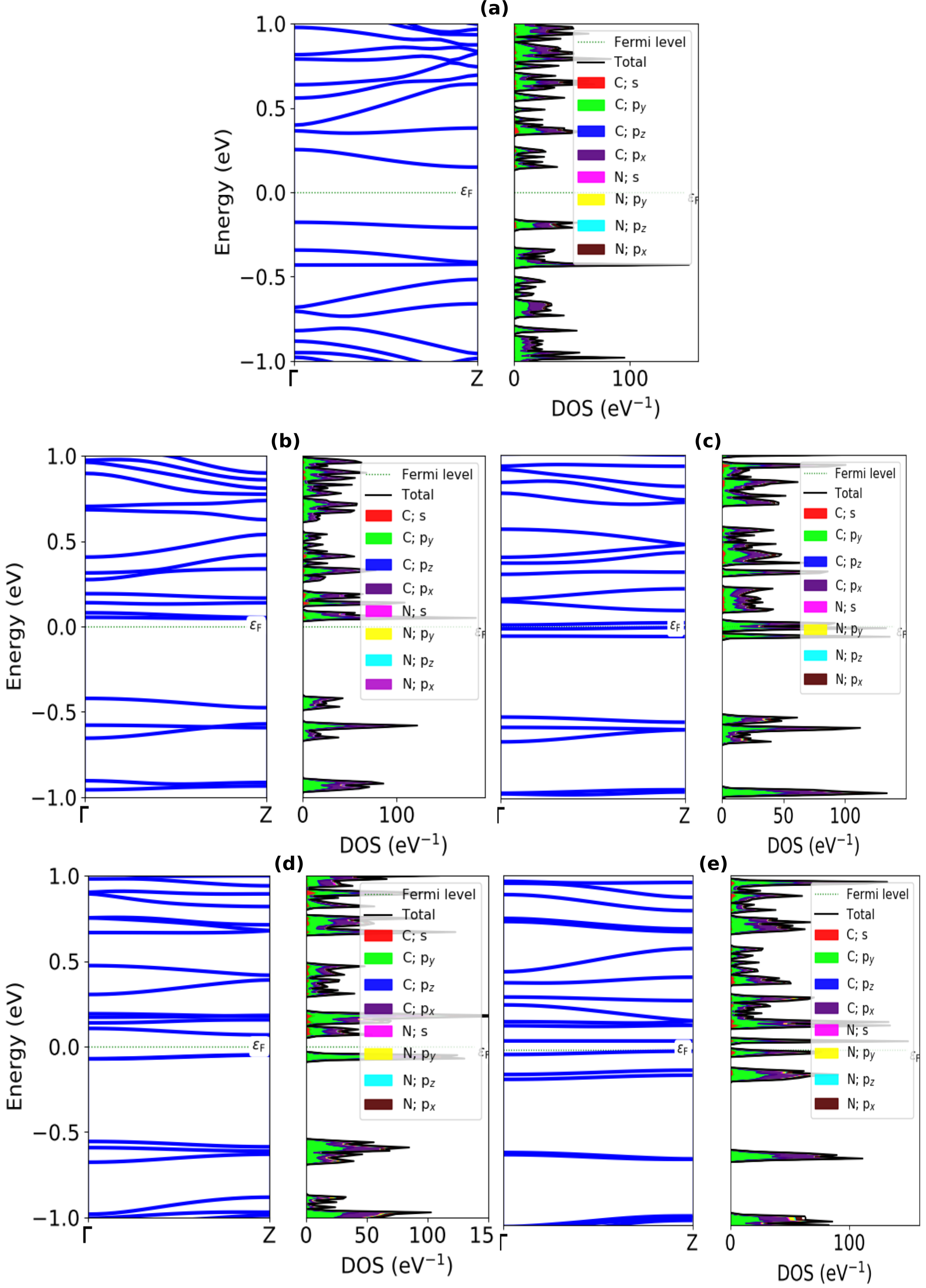}
	\caption{\label{fig:Band3}.Band structures and PDOS for (a) 2N system, (b) 6N system, (c) 10N system, (d) 12N system, (e) 14N system with zigzag doping }
\end{figure}
\begin{figure}[H]
	\centering
	\includegraphics[width=10.0cm, height=15.0cm]{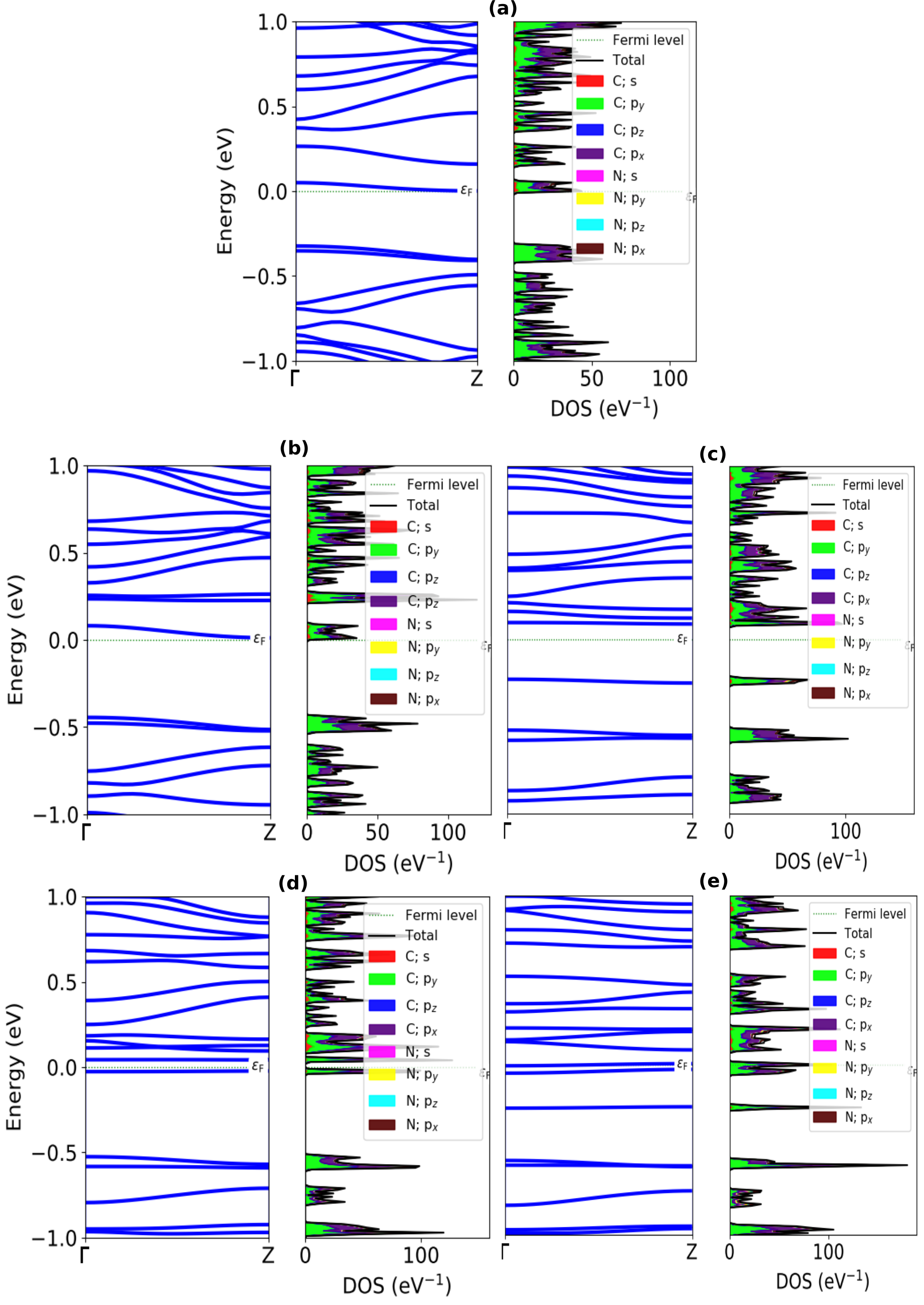}
	\caption{\label{fig:Band4}.Band structures and PDOS for: (a) 1N system, (b) 3N system, (c) 5N system, (d) 7N system, (e) 11N system with zigzag doping  }
\end{figure}
 
 For investigating the electronic properties of the N doped systems, we have calculated the band structure and projected density of states (PDOS). Figure (\ref{fig:Band1},\ref{fig:Band4}) shows the computed results. On comparing figure 7(a) and figure 8(a),two impurity states  are appeared in case of 1N-doped system, one is at the conduction band and another is at the valance band  reducing the band gap of the tube. From the PDOS (see figure 8(a) left), we find that these impurity states mostly consists of C-2p orbitals while the contribution of the N-2p orbitals are very less. Careful investigation show that the band gap  of the tube variate differently with even number ( 2N, 4N, 6N, etc.,) and odd number doping (1N,3N, 5N, 7N,  etc.) of N atoms as well as the doping pattern (arm-chair and zigzag doping  in our case). Similar results were also reported by  Allali et al. \cite{Allali2020a}. In odd number doping, the impurity state on the conduction band is appeared very closed to the Fermi level in comparison with that of the valence band (see figure \ref{fig:Band2} and figure \ref{fig:Band4}). The details of the investigation are shown in table \ref{tab:table1} and table \ref{tab:table2}.
 
\FloatBarrier
\begin{sidewaystable}[h]
	\centering
	\begin{tabular}{|c|c|c|c|c|c|c|c|}
		\hline
		 \multicolumn{4}{c|}{even doping} &
		\multicolumn{4}{c}{odd doping}\\
		\hline
		No. of N atoms & HOMO(eV) & LUMO(eV) & band gap(eV) & No. of N atoms& HOMO(ev) & LUMO(ev) & band gap(ev)\\
		\hline
		pristine & -0.247487 & 0.223758  & 0.471245 & 1N & -0.323106 & 0.00141296 & 0.27006 \\
		2N & -0.177499 & 0.148991 & 0.326491 &3N & -0.283293 & 0.0377814 & 0.321074 \\
		 4N& -0.100108 & 0.0830984 & 0.183207 & 5N & -0.18148 &0.0224506  & 0.204946 \\
		 6N& -0.052621 & 0.03053 & 0.0831510 & 7N & 0 & 0 & 0 \\
		 8N& -0.041676 & 0.0218104 & 0.06348 & 9N & -0.00890457 & 0.0360743 & 0.0449789 \\
		  10N& -0.0105665 & 0.00771054 & 0.0182771& 11N& -0.000923348 & 0.0665872 & 0.0675105 \\
		    12N& 0 & 0 & 0& 13N & 0 & 0 & 0 \\
		     14N& -0.0469492 & 0.0356408 & 0.08259&  &  &  &  \\
		      16N& -0.0761778 & 0.0048505 & 0.0806629& &  &  &  \\
		       18N& 0 & 0 & 0 & &  &  &  \\
		\hline
	\end{tabular}
	\caption{\label{tab:table1}Calculated electronic properties of arm-chair doping }
\end{sidewaystable}
\FloatBarrier
\FloatBarrier
\begin{sidewaystable}[h]
	\centering
	\begin{tabular}{|c|c|c|c|c|c|c|c|}
		\hline
		\multicolumn{4}{c|}{even doping} &
		\multicolumn{4}{c}{odd doping}\\
		\hline
		No. of N atoms & HOMO(eV) & LUMO(eV) & band gap(eV) & No. of N atoms& HOMO(ev) & LUMO(ev) & band gap(ev)\\
		\hline
		2N & -0.177472 & 0.148965 & 0.326437 &1N & -0.323094 & 0.00141646 & 0.324511 \\
		4N& -0.0602263 & 0.151255 & 0.211481 & 3N & -0.444425 & 0.0114775 & 0.455902 \\
		6N& -0.420965 & 0.0474502 & 0.468415 & 5N & -0.227647 & 0.0901583 & 0.317805 \\
		8N& -0.122244 & 0.0193635 & 0.141607 & 7N & -0.0227414 & 0.042464 & 0.0652054 \\
		10N& -0.00655758 & 0.00836462 & 0.0149222 & 9N& -0.324017 & 0.0125514 & 0.336569 \\
		12N& -0.484136 & 0.0682904 & 0.116704 & 11N & 0 & 0 & 0 \\
		14N& -0.00661522 & 0.051628 & 0.582432& 13N & -0.120072  & 0.00256388 & 0.122636 \\
		16N& -0.00929701 & 0.0985099 & 0.107807 &   &  &  &  \\
		18N& 0.113381 & 0.0849345 & 0.0962725 & &  &   &  \\
		\hline
	\end{tabular}
	\caption{\label{tab:table2}Calculated electronic properties of zigzag doping pattern}
\end{sidewaystable}
\FloatBarrier
 In  arm-chair doping pattern, when the nitrogen atoms are increased in  odd numbers, a little effect is appered on the conduction band. But in valence band, the highest energy states moves more closer towards the femi level  leding to the decrease in band gap. The first zero band gap value is obtained at 7N. For even doping  both the conduction and valence band are effected equally and the band gap decreses with incraese in N atoms with first zero band gap at 12N. Beyond 7N (for odd doping) and 12N  (for even doping) the band gap values oscillates. In case of zigzag doping , similar results  with that of the arm-chair doping pattern are observed for odd number doping. But the  zero band gap value is obtained at 11N. Beyond 11N the band gap values oscillates. When nitrogen atoms are doped in even, no zero band gap is observed and the band gape vlues oscillates with increase in impurity concentration. Jianhao et al.\cite{Shi2016a} and Zhao et al.\cite{Zhao2014a} also reported  similar results. On comparing the band gap values, pristine CNT has the largest band gap value. Our result is also in good agreement with Tetil et al. \cite{Tetik2014}.

\subsection{Mechanical properties}
\begin{figure}[H]
	\centering
	\includegraphics[width=10.0cm, height=5.0cm]{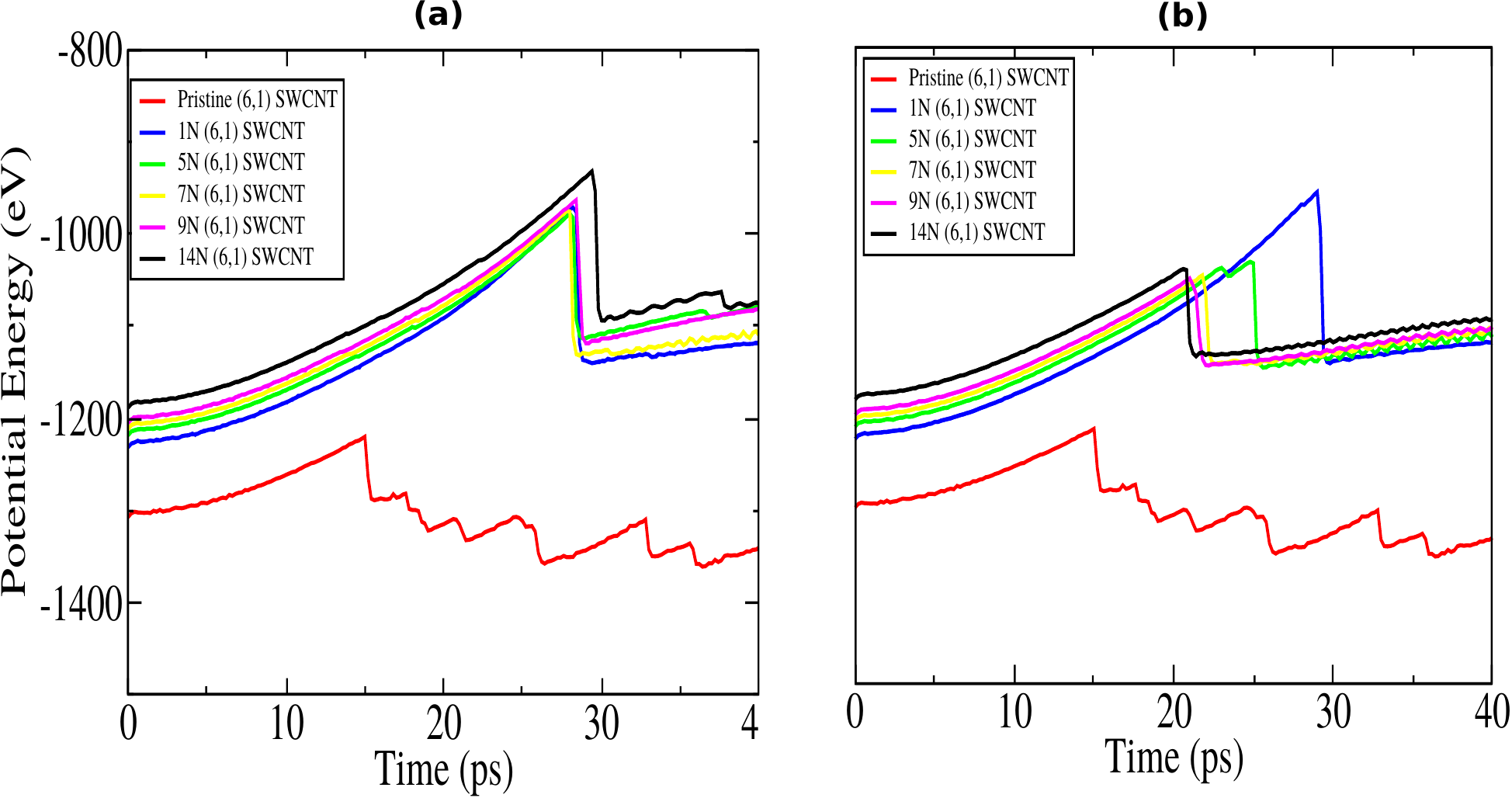}
	\caption{\label{fig:potnt}. (a) Potential-time period for arm-chair doping  (b) Potential-time period zigzag doping  }
\end{figure}

\begin{figure}[H]
	\centering
	\includegraphics[width=10.0cm, height=5.0cm]{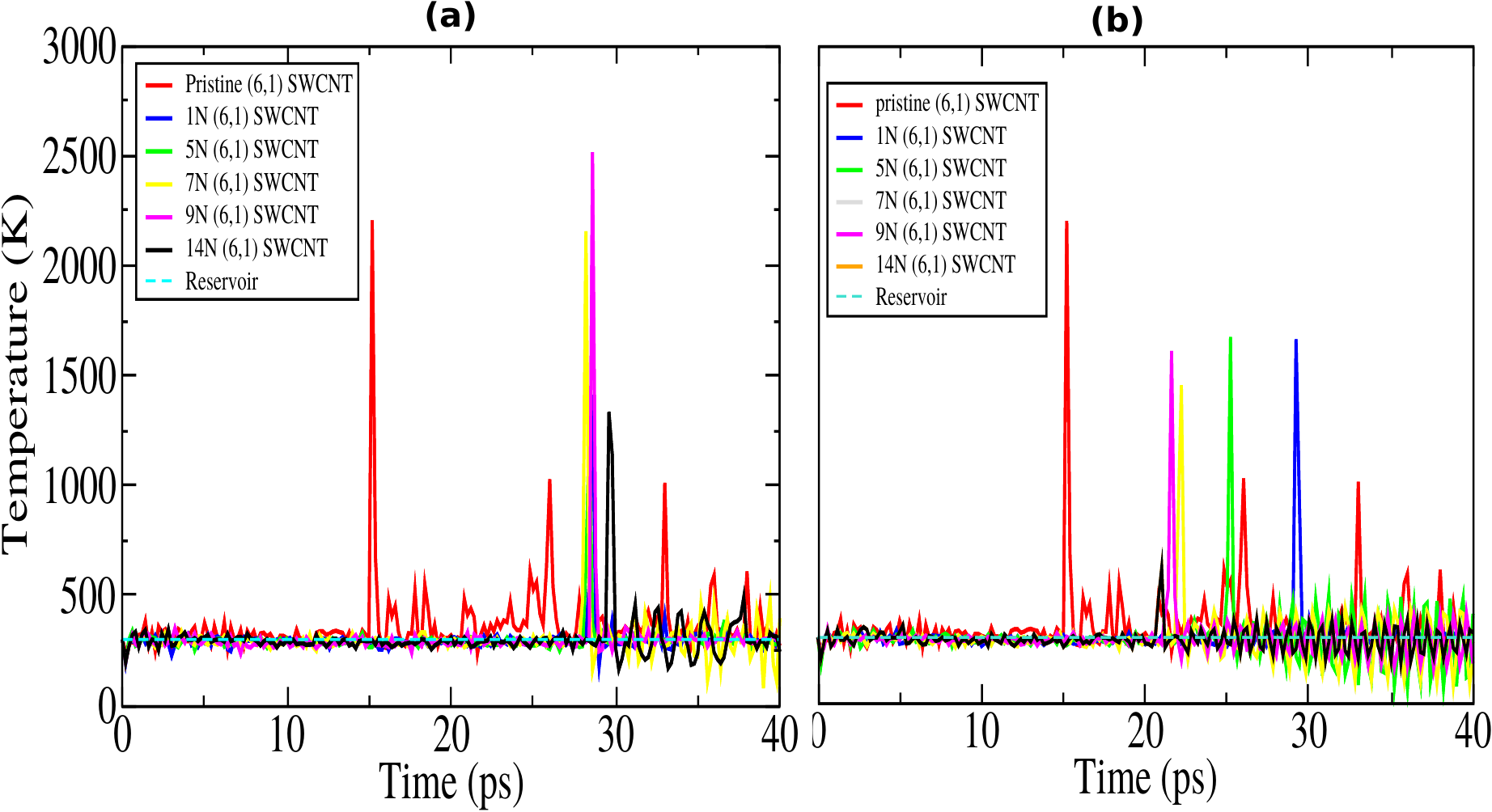}
	\caption{\label{fig:tem}. Temperature-time period for  (a) arm-chair doping  (b)Temperature-time period zig-zag doping .}
\end{figure}

 For the mechanical properties we performed a series of Molecular Dynamics (MD) simulation  The results of our calculations are shown in figure \ref{fig:potnt} and figure \ref{fig:tem}.  In temperature-time graph the peak point indicates the point of fracture of the CNT. Before the point of fractured, the temperature remain fluctuated at around 300k .The potential-time graph shows negative values thoughout the simulation process. These indicates that our systems are stable and also possible to synthesis at room temperature.

\begin{figure}[H]
\centering
\includegraphics[width=10.0cm, height=5.0cm]{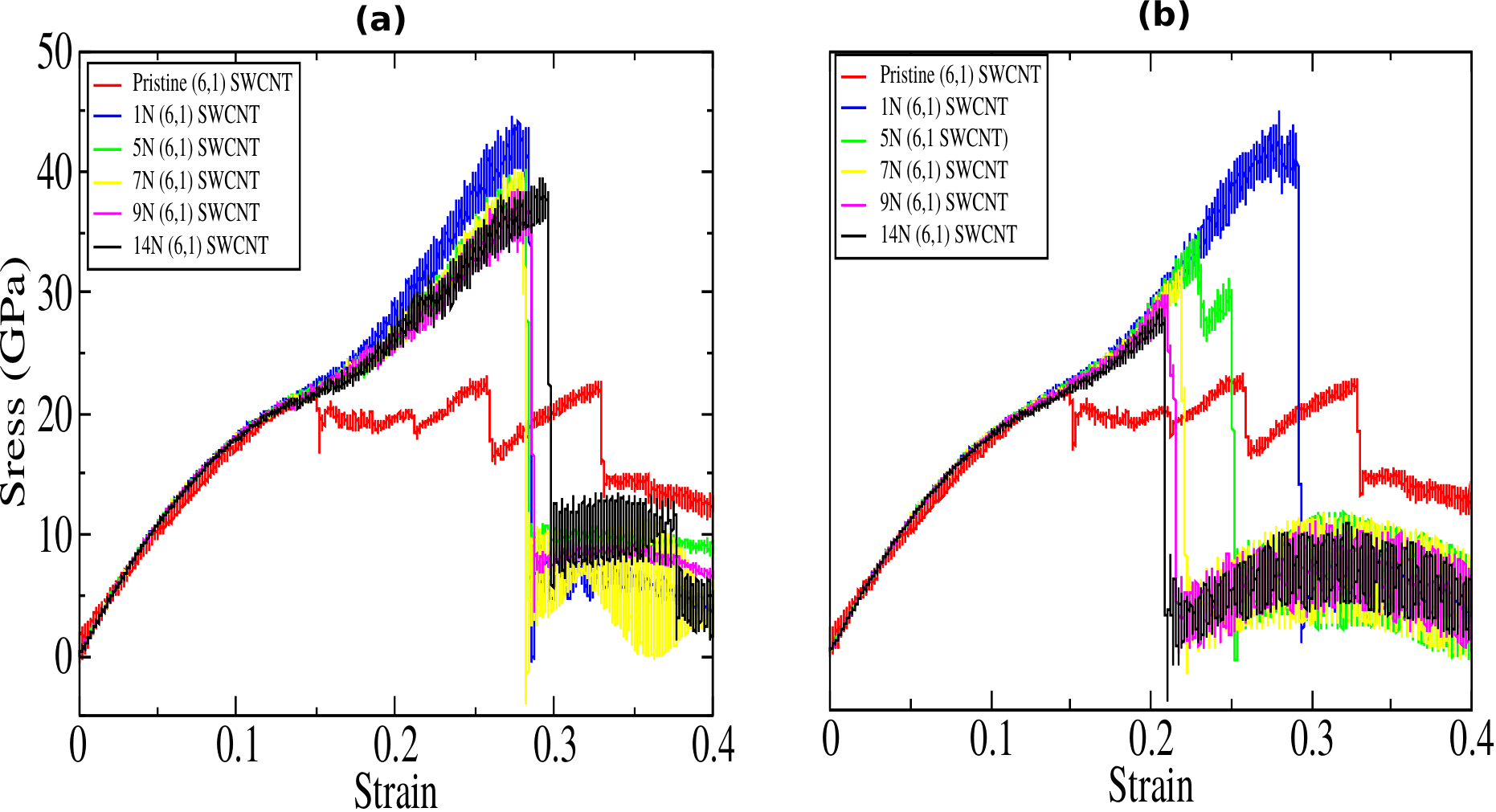}
\caption{\label{fig:strain}.Stress-Strain curve for (a) arm-chair doping  (b) zig-zag doping  }
\end{figure}

The stress for paristine (6,1) SWCNT increses when strain value is gradually raised  upto  15\%. Beyond strain value of 15\%, yield formation is observed and extended upto strain value of 35\%. Right after that, the stress value suddenly jump at nearly unchanged strain which indicates fracture in the SWCNT. Such a tensile response of the tube shows singnifiant change when doped with nitrogen atoms at different concentrations and doping patterns (see figure \ref{fig:strain}).In both zigzag and arm-chair doping , no yield formation are observed. 1N doped (6,1) SWCNT shows a maximum tensile stress value of 45 GPa which is 55\% higher than that of pristine one. But the strain at the point of fracture is slightly reduced from that of the pristine SWCNT. With arm chair doping , the increase in dopant concentration  shows insignificant effect in tensile response (see figure \ref{fig:strain}(a)). But in case of zigzag doping, the tesile strength and the critical strain value of the tube significantly reduce with increase in dopant concentration (see figure \ref{fig:strain}(b) ). 

\begin{figure}[H]
	\centering
	\includegraphics[width=10.0cm, height=5.0cm]{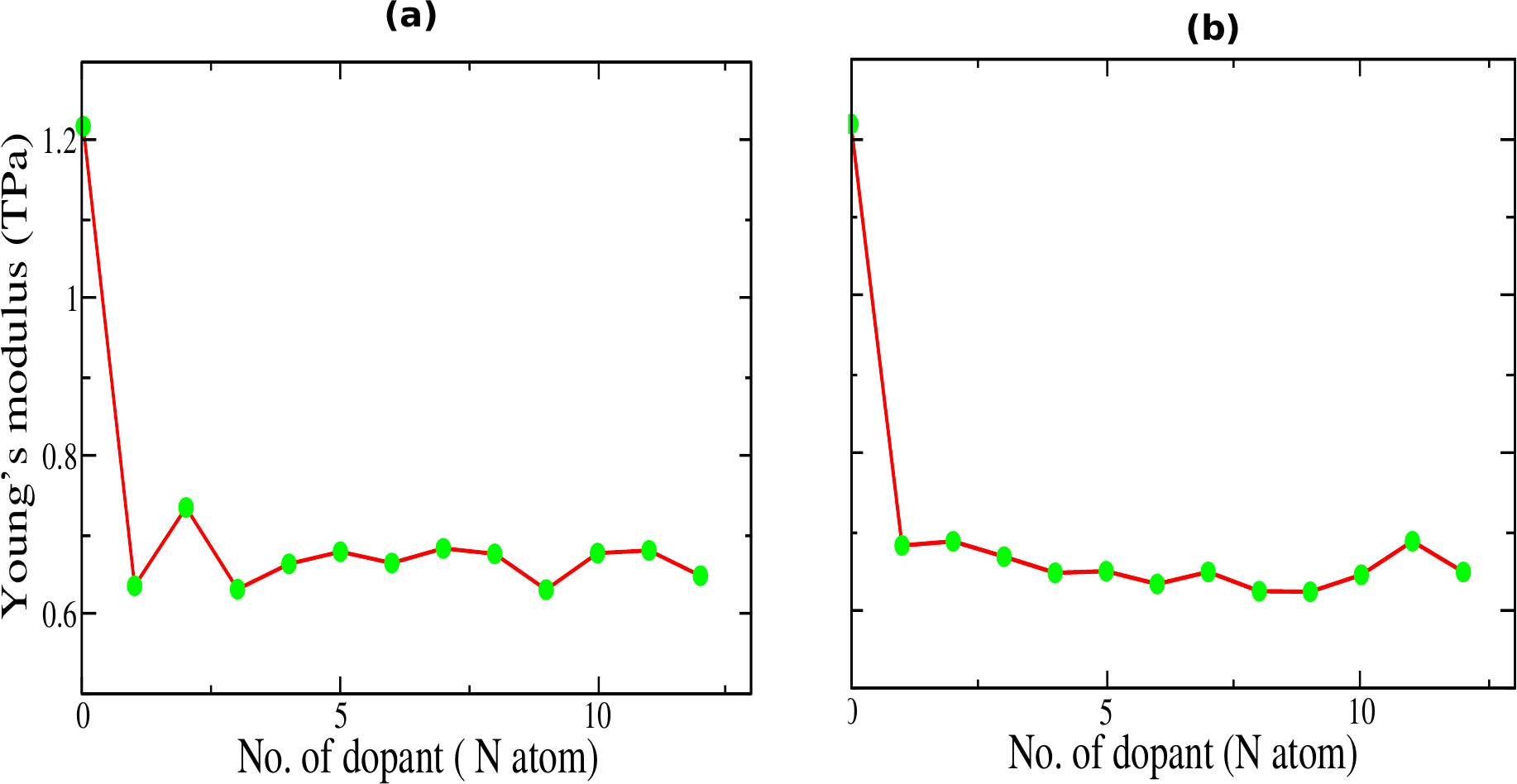}
	\caption{\label{fig:young}.Young's modulus and number of dopped N atoms curve for (a) arm-chair doping pattern (b) zig-zag doping pattern doped.}
\end{figure}
From strain and stress curve, the Young's modulus  for all the systems are  calculated. Figure \ref{fig:young} shows the calculated results. For pristine (6,1) SWCNT, the  Young's modulus values of 1.218 TPa is obtained which is in good agreement with previous investigation results \cite{Lei2011, Zang2009}. For 1N  system, the young's modulus value reduced to 0.735 TPa. Further increase in dopant concentration shows insignificant effect on Young's modulus value, also reported by Xia et al. \cite{Xia2014}.

\section{Conclusion}

A theoretical investigation based on  DFT about the effects of nitrogen doping on the electronic and mechanical properties 0f (6,1) SWCNT has been performed. The phonon DOS and formation energy calculation proves the stability of our systems. The electronic properties of the (6,1) SWCNT vary differently with the difference in doping pattern. In the same doping pattern, even number doping and odd number doping of the nitrogen atoms  also results difference in the electronic properties of the tube. The maximum energy band gap is obtained in the pristine CNT. The mechanical properties of the tube also shows significant effect on the doping concentration and doping pattern. One nitrogen doped (6,1) SWCNT shows the enhancement in tensile stress by 55\%. But the Young's modulus value is reduces to 0.735 TPa. Differ in the doping pattern and the doping concentration have very less effect in the Young's modulus value of the tube. We hope that our results may be helpful in practical engineering applications. 

\begin{acknowledgement}
\textbf{D. P. Rai} thanks Core Research Grant from Department of Science and Technology SERB (CRG DST-SERB, New Delhi India) via Sanction no.CRG/2018/000009(Ver-1).\\
\end{acknowledgement}


\bibliography{library}

\end{document}